\documentclass[aps,amssymb,article,pra,twocolumn,showpacs,showkeys,floatfix]{revtex4}

\usepackage{epsf}
\usepackage{amsmath,bm}
\usepackage{graphicx}
                      
\begin{document}
         
  \title{Radiation of anomalous energy}
  
  \author{B. Ivlev}
  
  \affiliation{Instituto de F\'{\i}sica, Universidad Aut\'onoma de San Luis Potos\'{\i},\\ 
  San Luis Potos\'{\i}, 78000 Mexico}

  \begin{abstract}
  
  In the Coulomb field of nucleus cut off on its size, besides usual atomic states, there are additional ones. These anomalous states are deep (in the range of $10\,MeV$) in the Dirac sea and can exist 
  solely during a macroscopic acceleration of the nucleus exceeding the certain level. This way a low energy perturbation, providing the nucleus acceleration, can result in $\sim 10\,MeV$ gamma and 
  neutron radiation due to electron transitions to the anomalous levels. This is anomalous but not nuclear energy. Experimentally observed gamma quanta and neutrons of that energy scale, in the 
  electric discharge in air, look as radiation from ``nothing'' if to forget about the transitions to the anomalous states. The perspective of energy generation, when the energy yield exceeds the input 
  level, is discussed. 
  
  \end{abstract} \vskip 1.0cm
  


  \maketitle
  
  \section{PARADOXICAL EXPERIMENTS}
  \label{exp}
  In experiments \cite{OGI1,OGI2} the high voltage discharge in air was revealed to produce the gamma and neutron radiation in the $10\,MeV$ range. This radiation penetrated through the
  $10\,cm$ thick lead wall. Within one discharge event the radiation elapsed approximately $10\,ns$ and corresponded to $10^{14}$ gamma quanta per second. 
  
  Since the applied voltage was not larger than $1\,MV$, the bremsstrahlung energy, like in an X-ray tube, could not exceed $1\,MeV$. Another source of the observed high energy radiation could be 
  nuclear reactions.  In \cite{BAB} the radiation was analyzed in details and it was concluded that ``known fundamental interactions cannot allow prescribing the observed events to neutrons''. The 
  authors of \cite{OGI1,OGI2} were surprised by the paradoxical inconsistency. 
  
  In experiments \cite{OGI1,OGI2} it was a small power station working during $10\,ns$ (per one discharge event) and generating $100\,W$ from ``nothing'' in the form of high energy radiation in 
  the $10\,MeV$ region.
  
  In Ref.~\cite{FOI1} the electric explosion of titanium foils in water resulted in changes of concentration of chemical elements. The applied voltage of $\sim 10\,keV$ could not accelerate ions
  in the condensed matter up to nuclear energies required for element transmutations. The surprising observations of neutrons from solids under mechanical perturbations were reported in \cite{DER,CAR}. 
  
  The phenomena in \cite{FOI1,DER,CAR} are also paradoxical since they are impossible without high energy processes. But how these processes could be caused by the relatively low applied voltage 
  in \cite{FOI1} or by the conventional (low energy) mechanical perturbations in \cite{DER,CAR}?
  
  \section{THE MECHANISM}
  \label{mech}
  The above experimental results can not be explained by known effects. A different concept is required, which provides a link between usual macroscopic phenomena and high energy processes. 
  
  It happens that in the Coulomb field of nucleus cut off on its size, besides usual atomic states, there are additional ones \cite{IVL}. These anomalous states are deep, in the range of $10\,MeV$,
  in the Dirac sea and can exist solely during a macroscopic acceleration (deceleration) of the nucleus. According to preliminary estimates, this acceleration should exceed the certain fluctuation 
  background. Note that this condition does not require a large kinetic energy of the atom.
  
  An example of that acceleration condition can be the enhancement of the atom velocity at least by $10^3m/s$ during $10^{-13}s$. This is not a conventional process. The necessary extreme but 
  reachable acceleration conditions \cite{IVL} were met in the experiments \cite{OGI1,OGI2,FOI1,DER,CAR} including fast processes in the lab lightning. Now one can look from the different angle at 
  ball lightning. Instead of an outer energy source, keeping the ball steady, it could be the internal (anomalous) energy source.   
  
\begin{figure}
  \includegraphics[width=7.0cm]{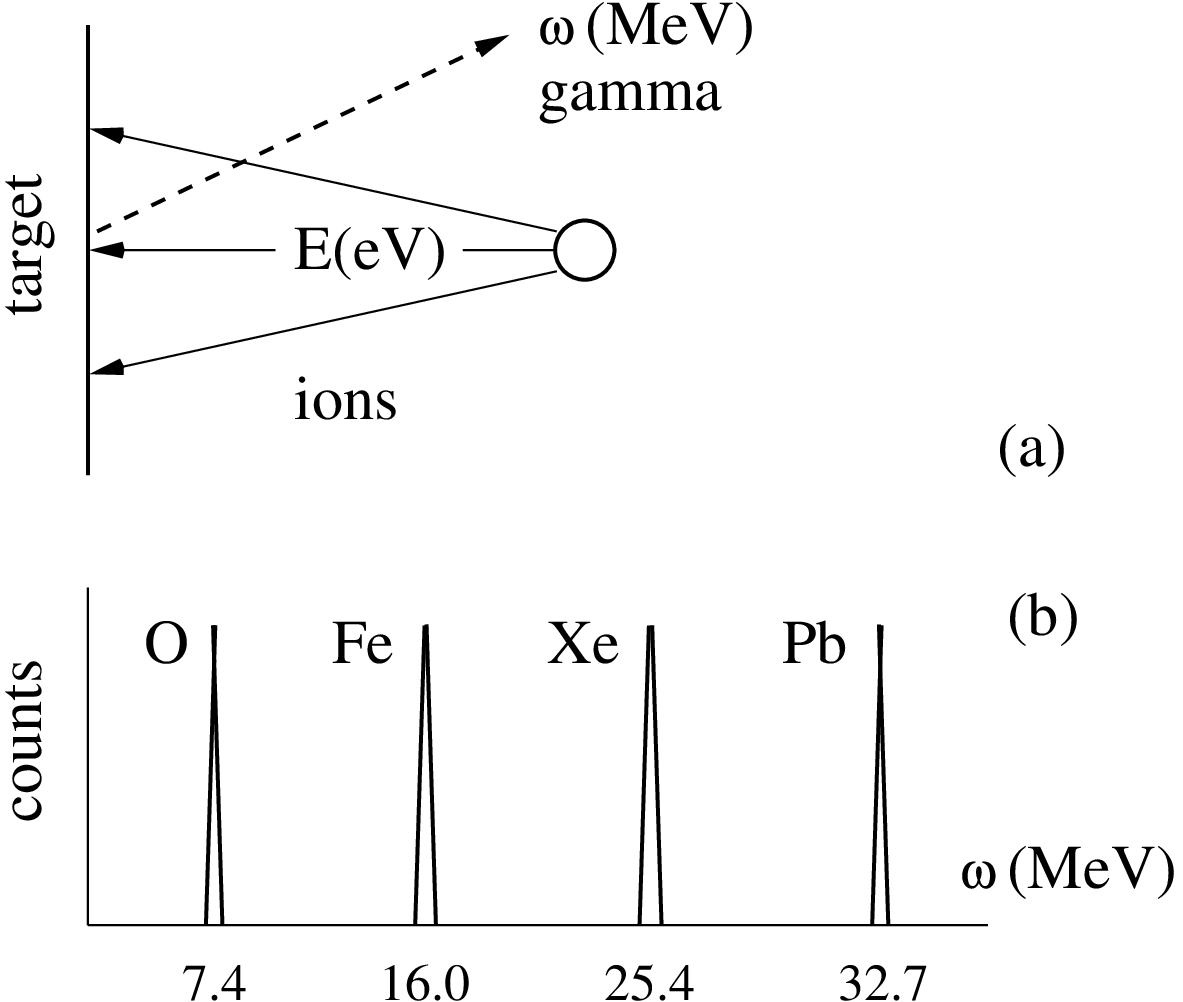}
  \caption{\label{fig1}(a) Low energy ion beam ($E\sim 100\,eV$) triggers off the radiation of the high energy $(\omega\sim 10\,MeV)$ quantum. (b) Anomalous gamma radiation from various nuclei. 
  Additional details are described in \cite{IVL}.}
  \end{figure}

  Returning to the anomalous processes, the electron falls to the anomalous state, from a usual atomic level, with the rate of $10^91/s$ during the entire period of atom acceleration producing the 
  gamma radiation with the spectrum shown in Fig.~\ref{fig1}(b). Those high energy electrons can also directly activate nuclear vibrations resulting in neutron emission like in fission \cite{IVL}. 
  These high energy processes can be referred to as radiation of anomalous energy \cite{IVL}. Despite it is in the $10\,MeV$ range this is not nuclear energy. The phenomenon corresponds to the 
  different aspect of high energy physics.  
  
  The anomalous mechanism is not within the set of known phenomena. It links two different worlds: usual low energy perturbations, providing the nucleus acceleration, and high energies as in nuclear 
  reactions. This bridge is compatible with the paradoxical experimental results \cite{OGI1,OGI2,FOI1,DER,CAR}. 
  
  \section{ION BEAMS}
  \label{beams}
  Ion beams are used to affect matter. See for example \cite{VAL}. There is a different aspect of the beam physics related to anomalous radiation \cite{IVL}. A steady ion beam, colliding a target, 
  is expected to permanently reproduce the high energy emission observed in \cite{OGI1,OGI2} solely $10\,ns$ in one discharge shot. This is because instead of one shot there are many small shots 
  corresponding to target collision by individual ions in the beam.
  
  Whereas in experiments \cite{OGI1,OGI2} the high energy from ``nothing'' was observed (pointing to formation of anomalous states), the analogous phenomenon in ion beams was not reported in 
  literature. This opens a way to study a possibility of high energy radiation produced by low energy ion beams. That radiation would look as coming from ``nothing'' if to forget about the transitions 
  to the anomalous states. 
  
  Formation of anomalous states (and thus the energy generation) in an ion beam can occur under its collision with a target. The conditions for that is a sufficiently large deceleration of ions by the 
  target. This depends on ion energy and a type of the target since the ions can be stopped quickly or penetrate deep in the target. 
  
  An ion in Fig.~\ref{fig1}(a), even of the low energy preliminary estimated as $E\sim 100\,eV$, after collision a proper chosen target can trigger off the anomalous gamma radiation in the region 
  of $10\,MeV$. The peak positions of the gamma radiation in Fig.~\ref{fig1}(b) do not depend on ion energy $E$, which can be in a wide region. The width of each peak is approximately $0.1\%$ of 
  its position. 
  
  With the proper fitting of the target and the $10\,kA$ beam, one can expect the energy yield up to $1\,MW$ in the form of gamma radiation in the $10\,MeV$ range. 
  
  ~
  
  \section{PERSPECTIVE OF ENERGY GENERATION}
  \label{ene}
  According to evaluations in \cite{IVL}, in the experiments \cite{OGI1,OGI2} $10^{5}$ ions were in the anomalous state (that is with the proper acceleration) every moment of time during $10\,ns$
  generating $100\,W$ in the form of high energy, $\sim 10\,MeV$, radiation. 
  
  Suppose that in the certain process $10^{18}$ atoms (approximately $ 0.1\,mg$ of matter) are properly accelerated. Creation of such extreme conditions is a matter of study. That power station will 
  produce $10^{15}W$ during the acceleration process. For comparison, the power of that station would exceed two orders of magnitude the power of the biggest industrial power plant. The energy of 
  $10^{15}J$ is released under the explosion of one megaton of trotyl.
  
  \section{CONCLUSIONS}
  \label{con}
  When an atom is accelerated, besides bremsstrahlung the high energy ($\sim10\,MeV$) gamma and neutron emission is possible during the acceleration. This is not nuclear but the electron process of 
  transitions to deep lying (anomalous) electron levels formed during the acceleration. The criterion of this anomalous process is not the kinetic energy, acquired by the atom, but its sufficiently 
  large acceleration. For example, the atom can be accelerated up to $10^3m/s$ during $10^{-13}s$ acquiring thus the kinetic energy $\sim 1\,eV$. This is an extreme but realizable process. 
  
  It is highly likely that the paradoxical radiation of high energy quanta and neutrons, observed in the electric discharge in air \cite{OGI1,OGI2}, was caused by electron transitions to the anomalous 
  levels. During $10\,ns$ this small power station produced $100\,W$ in the form of $\sim 10\,MeV$ radiation. It was anomalous but not nuclear energy. This can be a prototype of real devices based on 
  anomalous radiation. The phenomenon corresponds to the different aspect of high energy physics.

  \end{document}